\documentclass[conference]{IEEEtran}
\IEEEoverridecommandlockouts
\usepackage{cite}
\usepackage{amsmath,amssymb,amsfonts}
\usepackage{algorithmic}
\usepackage{graphicx}
\usepackage{subfigure}
\usepackage[algo2e,linesnumbered,ruled]{algorithm2e}

\def\BibTeX{{\rm B\kern-.05em{\sc i\kern-.025em b}\kern-.08em
    T\kern-.1667em\lower.7ex\hbox{E}\kern-.125emX}}

\begin{document}

\title{Communications and Networking Technologies for Intelligent  Drone Cruisers
\thanks{L.-C. Wang, C.-C. Lai, and H.-H. Shuai are with the  Department of Electrical and Computer Engineering, National Chiao Tung University, Hsinchu, Taiwan. E-mail: lichun@cc.nctu.edu.tw. (Corresponding author: L.-C. Wang)}
\thanks{H.-P. Lin is with the Department of Electronic Engineering, National Taipei University of Technology, Taipei, Taiwan.}
\thanks{C.-Y. Li is with the Department of Computer Science, National Chiao Tung University, Hsinchu, Taiwan.}
\thanks{C.-H. Chen and T.-H. Cheng are with the Department of Mechanical Engineering, National Chiao Tung University, Hsinchu, Taiwan.}
\thanks{This research is supported by Ministry of Science and Technology under the Grant MOST 108-2634-F-009-006- through Pervasive Artificial Intelligence Research (PAIR) Labs, Taiwan.}%
}

\IEEEspecialpapernotice{(Invited Paper)}

\author{\IEEEauthorblockN{Li-Chun Wang, Chuan-Chi Lai, Hong-Han Shuai, Hsin-Piao Lin, Chi-Yu Li, Teng-Hu Cheng, \\ and Chiun-Hsun Chen}%
}

\maketitle

\begin{abstract}
Future mobile communication networks require an Aerial Base Station (ABS) with fast mobility and long-term hovering capabilities. At present, unmanned aerial vehicles (UAV) or drones do not have long flight times and are mainly used for monitoring, surveillance, and image post-processing. On the other hand, the traditional airship is too large and not easy to take off and land. Therefore, we propose to develop an ``Artificial Intelligence (AI) Drone-Cruiser” base station that can help 5G mobile communication systems and beyond quickly recover the network after a disaster and handle the instant communications by the flash crowd. The drone-cruiser base station can overcome the communications problem for three types of flash crowds, such as in stadiums, parades, and large plaza so that an appropriate number of aerial base stations can be accurately deployed to meet large and dynamic traffic demands. Artificial intelligence can solve these problems by analyzing the collected data, and then adjust the system parameters in the framework of Self-Organizing Network (SON) to achieve the goals of self-configuration, self-optimization, and self-healing. With the help of AI technologies, 5G networks can become  more intelligent. This paper aims to provide a new type of service, On-Demand Aerial Base Station as a Service.  This work needs to overcome the following five technical challenges: innovative design of drone-cruisers for the long-time hovering, crowd estimation and prediction, rapid 3D wireless channel learning and modeling, 3D placement of aerial base stations and the integration of WiFi front-haul and millimeter wave/WiGig back-haul networks.
\end{abstract}


\section{Introduction}
\label{intro}
\emph{Unmanned aerial vehicles} (UAVs) are becoming more and more popular due to their high mobility and low cost. 3GPP began discussing the integration of drones into the new 5G LTE-A standard in early 2018, but still regards drones as mobile three-dimensional terminals.  
The drone can be used not only as a device for surveillance applications but also as a flying base station (BS) that dynamically supports communications services. 
Since UAV base stations can be deployed in temporary locations and removed, UAV-BS is a highly mobile and economical solution that provides communication services for a large number of temporarily crowded people, such as concerts, baseball games, and emergencies.

Compared to terrestrial base stations deployed in a two-dimensional plane, the three-dimensional space makes the UAV-BS more flexible in deployment, regardless of geographic location. However, there are still many unresolved challenges and issues that need to be addressed when deploying UAV-BS in cooperation with terrestrial base stations to provide communications services. The challenges are briefly described below.

\begin{figure*}[!t]
	\centering
	\includegraphics[width=0.7\textwidth]{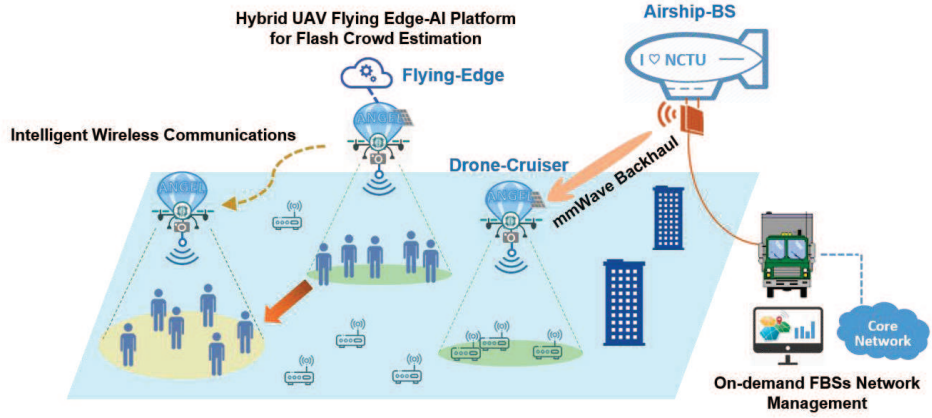}
	\vspace{-5pt}
	\caption{The new architecture of cellular system applied AI technologies/data sciences can support the on-demand UAV-BS as a service.}
	\label{fig:system}
	\vspace{-10pt}
\end{figure*}
\begin{itemize}
\item Crowd estimation and mobility prediction: In order to make the communications services intelligent, the system needs to know the communications demand triggered by spatio-temporal hot-spots. Therefore, we have a camera on the UAV-BS that integrates computer vision technology (including crowd counting and estimation) into the UAV-BS and helps predict communications demands. In addition, UE mobility prediction/management is also an important issue in improving communication performance and providing seamless services. 
\item 3D channel state learning: The UAV-BS deployment location has a large impact on the overall mobile communication network performance. In a dynamic communications network, the link quality between the user and the UAV-BS will vary with the wireless channel, user mobility, communication scenarios, and the location and altitude of the UAV-BS. The key to finding the effective location and height of the UAV-BS with high communications quality is the ability to quickly learn and identify the characteristics of the current 3D wireless channel ~\cite{6863654}~\cite{7881122}. 
\item On-demand 3D placement of UAV-BS: If the 3D channel, communication requirements (crowd estimation), and UE mobility prediction/information are available, the system can provide on-demand 3D placement recommendations for UAV-BSs. Therefore, the fusion of multi-source cross-domain information for 3D placement recommendation of UAV-BSs is also an indispensable research topic.
\item Robust and high-throughput backhaul: The UAV-BS needs to serve a large number of UEs, so the data rate and throughput of the backhaul connection is an important influence on system performance.
Therefore, how to ensure a robust backhaul connection between the UAV-BS and the ground base station is also a challenge.
To solve such a problem, there are two promising solutions: (1) Tethered UAV-BS or (2) millimeter wave (mmWave) communication technology. The formal one can satisfy the requirements of high throughput and long hovering time but make the deployment of the UAV base station inelastic. The latter one is an emerging wireless technology for high data throughput communication and it is more flexible to UAV communications.
\item Long-hovering time: Due to the limitation of the battery capacity of the UAV-BS and the design of the current mainstream UAV, the UAV cannot provide services for a long time. Therefore, the innovative design of a long-time hovering UAV-BS has become an important issue.
\end{itemize}


The balance of this paper is organized as follows. Section~\ref{sec:system_model_topics} presents the considered system model and the current research topics. Section~\ref{sec:recent_results} introduces the our recent proposed solutions with some preliminary results and identify some future directions. Finally, we conclude this paper in Section~\ref{sec:conclusion}.

\section{System Model and Research Topics}
\label{sec:system_model_topics}
In this paper, we thus design a management framework of UAV-BSs for emerging communications network applications. The considered system architecture of this work is shown in Fig.~\ref{fig:system}. In the existing wireless communication systems, artificial intelligence is often used for traffic and network monitoring and make the cellular system can automatically handle the problems of communication networks. For example, with AI technologies, the system can use the current state to determine how to deploy multiple UAV-BSs for effectively handling network offloading, relay transmission, and network load balancing. Such an innovative idea, UAV-BS, can support the existing wireless cellular systems. In fact, there are many relevant communication technologies that require further research and development. 

We aim at the communication network environment constructed by the UAV-BS. The key technology is to propose a UAV-BS management architecture and open up opportunities for innovative services of the mobile communications industry and the UAV manufacturing industry. Firstly, we analyze the characteristics of the 3D air-to-ground wireless channel and design a new learning mechanism for UAV-BS to provide stable communication services. We also consider some important characteristics, such as the 3D space characteristics of the deployment environment, network load, and signal interference, for optimizing the configuration and placement of the UAV-BS. Finally, we consider an issue, lifetime, of the UAV-BS. To solve this issue, we not only consider optimizing the power consumption from the communications aspect but also design a new type of UAV-BS which can hover for a long time.

To develop and implement the proposed system in Fig.~\ref{fig:system}, we suggest and conduct the following five research topics: (1) crowd estimation for communication demand prediction, (5) 3D channel modeling for link quality prediction, (3) flexible 3D placement of UAV-BSs, and (4) millimeter wave for UAV back-haul communications, and (5) the design of UAV for long-time hovering. 

\subsection{Crowd Estimation for Communication Demand Prediction}
We also use computer vision with AI techniques to estimate the population density and distribution from the sensed image of the crowd. With the information of crowds, the UAV-assisted cellular system can optimize itself more dynamically and appropriately and thus improve the system performance. Computer vision has become one of the popular technologies for many UAV applications, such as security monitoring, disaster management, public space design, and intelligence gathering. However, there are still many challenges in image recognition for outdoor applications, such as shadows, irregular object distribution, scale changes, and different perspectives and complex backgrounds.
In fact, such a crowd counting problem has been discussed by many works for some traditional monitoring applications. Zhang et al.~\cite{7780439} proposed a training method based on a multi-column architecture to obtain different densities by using differently sized kernels in the network. Sindagi and Patel~\cite{8237468} proposed the \emph{Contextual Pyramid Convolutional Neural Network} (Contextual Pyramid CNN) for extracting global and local context information through the CNN. These obtained context information can achieve lower counting errors and thus improves the quality of the density map. However, the authors use multiple models to extract these global and local context features simultaneously, and the extraction of these features leads to increased computational costs and complex training processes. In addition, most existing methods only produce low-resolution density maps and it will impose some limitations in practical applications.

\subsection{3D Channel Modeling for Link Quality Prediction}
In recent years, with the rapid development of AI, some applications have applied machine learning (ML) algorithms in the field of wireless communications. For example, some popular clustering techniques and tracking algorithms were introduced in~\cite{8284053} and applied to wireless channel clustering and modeling. Esrafilian and Gesbert~\cite{8254657} proposed a method to classify the line-of-sight (LoS) and the non-line-of-sight (NLoS) channel conditions of the air-to-ground (A2G) communication link using machine learning algorithms, and further converted it into 3D city map information using building position and height reconstruction algorithms. Cheng et al.~\cite{6963485} used machine learning algorithms to solve the problem of indoor positioning and NLOS channel identification. However, the changes in the 3D outdoor environment are more complicated and dynamic than the indoor environment. Therefore, a set of wireless channel identification and modeling techniques suitable for the UAV-BS needs to be studied.

\subsection{Flexible 3D Placement of UAV-BSs}
Al-Hourani et al.~\cite{6863654} firstly discussed the relation between the altitude of a UAV-BS and its optimal coverage. They proposed an air-to-ground (ATG) channel model with derivations of the probabilities of LoS and NLoS links and their proposed channel model has been widely used in UAV communications. Kalantari et al.~\cite{7881122} proposed a backhaul-aware robust 3D placement of UAVBS to temporarily increase the network capacity or coverage of an area in 5G+ environments. Lyu et al.~\cite{7762053} provided a polynomial-time method to deploy multiple UAV-BSs at appropriate locations and minimize the number of UAV-BSs while considering the various densities of users. Wang et al.~\cite{8301585} considered the case that all users are uniformly distributed in a serving and provided a method for adjusting the required transmit power of the UAV-BS to cover all users. Most existing works focused on how to let deployed UAB-BSs provide the maximum coverage or maximize the number of served users. However, in practice, such a maximum coverage problem had been solved by some widely used solutions, such as ultra-dense small cells and Wi-Fi access points. Hence, UAV-BS communication is a more suitable solution for the dynamic demands (or hot-spots) which lead by flash crowds which is arbitrarily distributed.

\subsection{Millimeter Wave for UAV Back-haul Communications}
This topic is aimed at the integration of millimeter-wave technologies for drones. It mainly consists of two directions. The first one is that millimeter-wave communication may be used as a backhaul connection between the UAV-BS and the terrestrial base station. It can be known from [9] that the data rate will be as high as Gbps, and the frequency of the millimeter wave will have a fast attenuation characteristic. The millimeter-wave communication technology not only can support Gigabit wireless transmission rate which satisfies the needs of a base station but also increases the mobility and scalability of the UAV-assist cellular system. Each UAV-BS can support users to access the Internet via WiFi or LTE relay. The second direction is that mmWave-assisted localization. Based on the light resistance and low interference of mmWave, it is expected to achieve an accuracy of less than one-meter error for UAV localization. When a UAV returns to the charging area, the mmWave can be used for positioning and navigation, and parked at a predetermined position. Hence, the development of millimeter-wave communication technology for Gbps transmission and the millimeter-wave array beam with AI to accurately track the location of UAV in 3D space have become popular topics recently.

\subsection{The Design of UAV for Long-Time Hovering}
\begin{figure}[!t]
	\centering
	\subfigure[]{
		\label{fig:drone:airship} 
		\includegraphics[width=0.277 \textwidth]{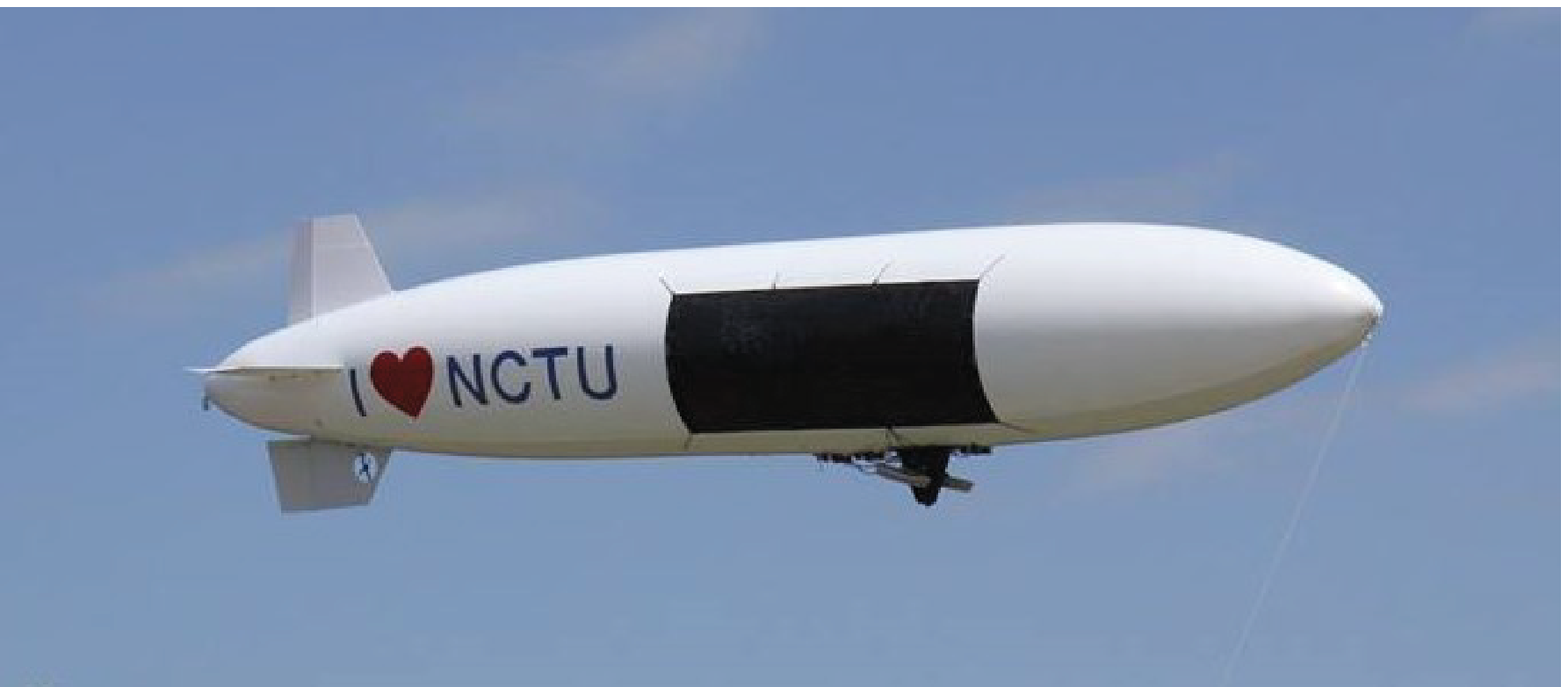}}%
	\subfigure[]{
		\label{fig:drone:commercial} 
		\includegraphics[width=0.193 \textwidth]{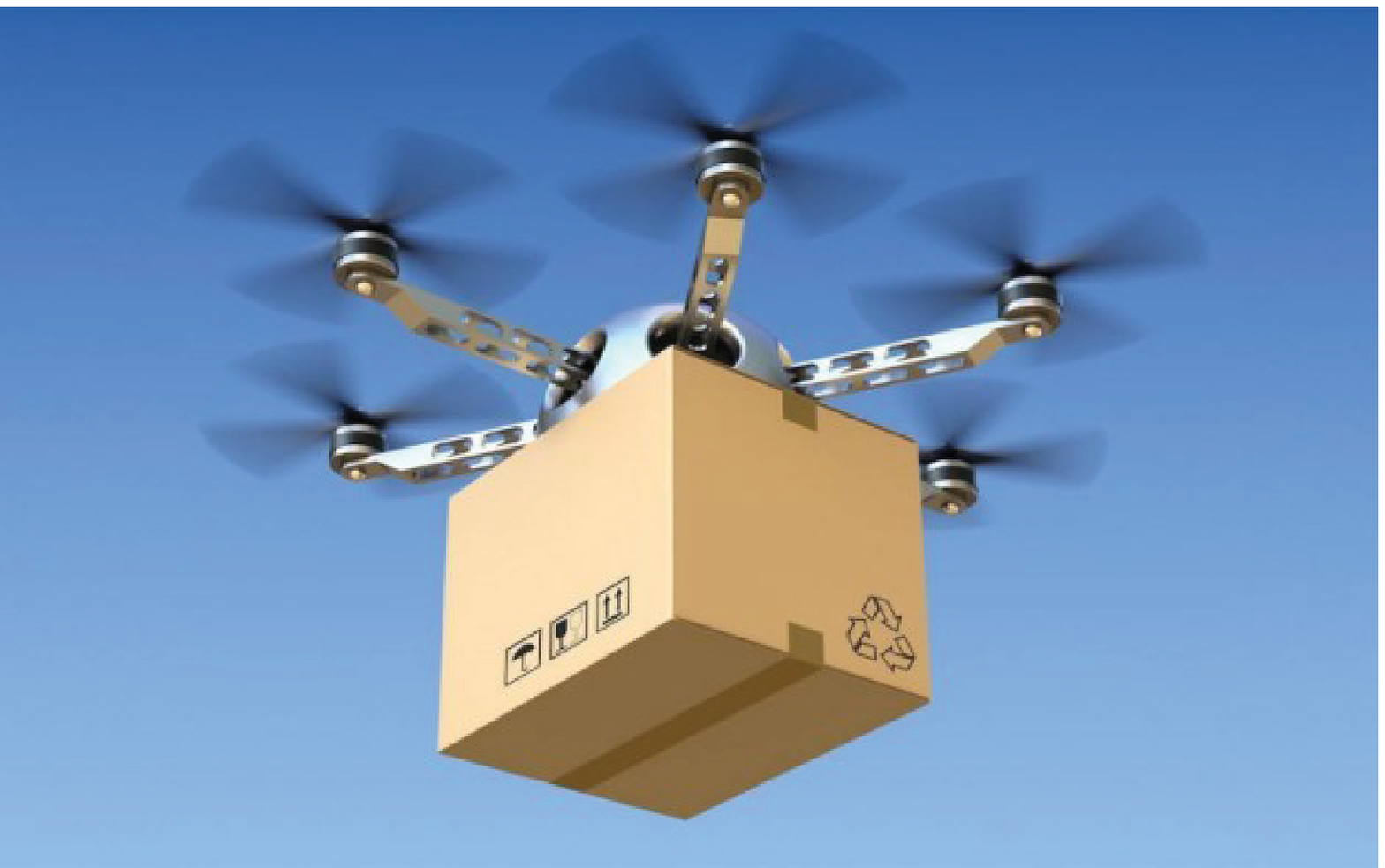}}%
	\caption{Two mainstream types of UAVs:~\subref{fig:drone:airship} large airships and ~\subref{fig:drone:commercial} commercial drones.
	}
	\label{fig:drone} 
	\vspace{-15pt}
\end{figure}

In order to monitor a large area, large airships, as shown as Fig.~\ref{fig:drone:airship}, are often used as vehicles, which have the advantage of being able to hang for a long time. However, for large airships, the relative volume is relatively large, the action is slow, and it is susceptible to external wind interference. In words, airships may not work due to the weather factor. Since an airship can stay in the air for a long time, it can also be used as a mobile flying base station (or relay station) for wireless communication. One possible application is using an airship mounted base station to provide wireless communication to the crowds and hotspots, so that many people can avoid local network congestion. However, such an application requires good flexibility and mobility to avoid the inability to fly due to external weather factors. Another solution is to use current commercial drones, as shown in Fig.~\ref{fig:drone:commercial}, which are highly maneuverable and have a small size. However, the capability of hovering for these commercial drones is limited by the poor power. According to the above reasons, we will design a new UAV-BS that combines the maneuvering capabilities of the drone with the advantages of airships.



\section{Recent Results and Future Directions}
\label{sec:recent_results}
After introducing the considered five research topics, we will detailedly address the proposed solutions for the above corresponding topics in following subsections.

\subsection{Crowd Estimation and Mobility Prediction}
\begin{figure}[!t]
	\centering
	\includegraphics[width=0.48\textwidth]{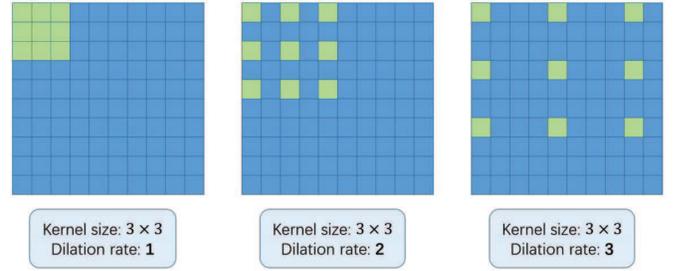}
	\caption{The cases of 3 $\times$ 3 convolution kernels using different dilation rate.}
	\label{fig:dilated_cnn}
	\vspace{-10pt}
\end{figure}

\begin{figure}[!t]
	\centering
	\subfigure[]{
		\label{fig:crowd_prediction:input} 
		\includegraphics[width=0.25 \textwidth]{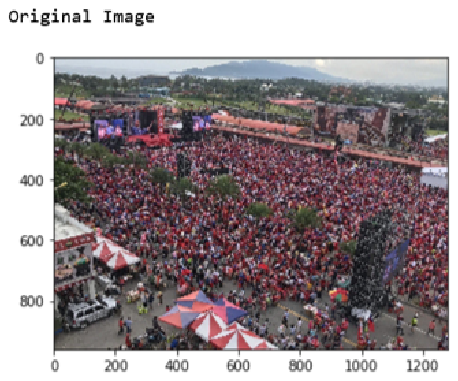}}%
	\subfigure[]{
		\label{fig:crowd_prediction:density_map} 
		\includegraphics[width=0.25 \textwidth]{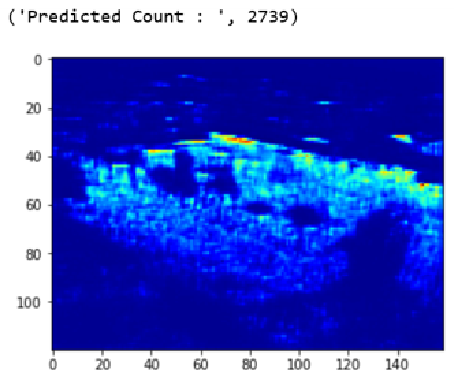}}%
	\caption{The result of crowd estimation:~\subref{fig:crowd_prediction:input} input image and ~\subref{fig:crowd_prediction:density_map} the output density map.
	}
	\label{fig:crowd_prediction} 
\end{figure}

In order to construct a model that can simultaneously perform crowd counting and density estimation tasks, we propose a framework created by a combination of a \emph{Dilated CNN}~\cite{YU2016ICLR} and a \emph{Long Short Term Memory Network} (LSTM). As shown in Fig.~\ref{fig:dilated_cnn}, dilated CNN uses sparse kernels to alternate the pooling and convolutional layer. The convolution kernels with different dilation rates are used to aggregate the multi-scale contextual information while keeping the same resolution. With the dilated CNN, the proposed framework can perceive a larger scale information and then conduct more precise estimations. In addition, each UAV-BS has to provide services for a long time, so we needs to know the continuous changes of the crowds which can help the decision-making of UAV-BS placement. Since LSTM is a \emph{Recurrent Neural Networks} (RNN) which can memorizes the previous states, it is suitable for handling time series (or sequential) data. Hence, we integrate a LSTM into our framework for predicting the estimation of crowds within a coming serving time period. Fig.~\ref{fig:crowd_prediction} presents the crowd estimation result of the proposed framework. As shown in Fig.~\ref{fig:crowd_prediction:input}, a lots of people in red and a large number of small red stools are in the input image. Fig.~\ref{fig:crowd_prediction:density_map} indicates that the proposed framework can correctly recognize the people who dress in red and the estimated count of people is 2739.

\begin{figure}[!t]
	\centering
	\includegraphics[width=0.8\columnwidth]{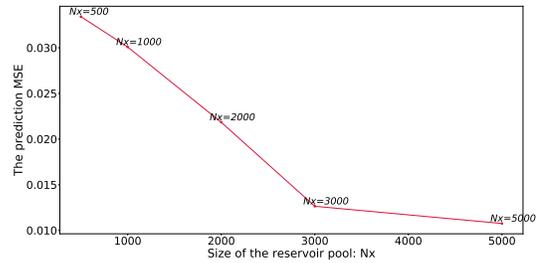}
	\caption{The performance of UE trajectory prediction in terms of RMSE~\cite{Peng1908:Predictive}.}
	\label{fig:esn_prediction}
	\vspace{-10pt}
\end{figure}

To provide seamless service, the system needs to handle the issue of UE mobility. We have propose a learning approach~\cite{Peng1908:Predictive} using \emph{Echo State Network} (ESN)~\cite{2001_Jaeger} for UE trajectory prediction. In this part, we use an open dataset, GeoLife~\cite{DBLP:journals/debu/ZhengXM10}, to train and validate the ESN model. Fig.~\ref{fig:esn_prediction} indicates that the proposed ESN-based learning approach can high quality of prediction and the Root Mean Square Error (RMSE) achieve 1.1\%. The RMSE is defined as 
\begin{equation}\label{eq:rmse}
RMSE=\sqrt{\dfrac{1}{n}\sum_{i=1}^n\left(C_i-\hat{C}_i\right)^2 },\nonumber
\end{equation}
where $n$ is the number of test images, the predicted count of image $i$ is $\hat{C}_i$ and the real count is $C_i$.

\subsection{3D Channel Fast Learning and Modeling}
\begin{figure}[!t]
	\centering
	\includegraphics[width=0.43\textwidth]{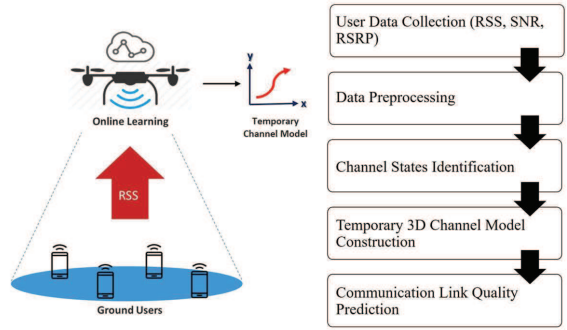}
	\vspace{-5pt}
	\caption{The flow chart of the online 3D wireless channel learning process~\cite{8651718}.}
	\label{fig:channel_learning_process}
	\vspace{-5pt}
\end{figure}

For this issue, we have proposed a 3D wireless channel fast learning and modeling technology~\cite{8651718} for dynamically learning the current status and extracting key features of the 3D wireless channel. With the proposed approach, the system can dynamically predict the approximate characteristics of the current communication environment and then provide a more precise prediction on the quality of service for the served users. As shown in Fig.~\ref{fig:channel_learning_process}, the proposed algorithm combines online learning with edge computing to reduces the computational complexity and thus achieves near real-time computation (low latency). The proposed algorithm can quickly analyze the characteristics of the air-to-ground 3D wireless channel and key parameters by using the current RSS information. After that, the system uses the unsupervised learning clustering method to establish a temporary wireless channel model which is more suitable for the current communication scene/scenario. In short, the proposed algorithm can be divided into the following five steps: 1) User Data Collection; 2) Data Pre-processing; 3) Channel States Identification; 4) Temporary 3D Channel Model Construction; and 5) Communication Link Quality Prediction. With the above process, the system can be combined adaptive communication with UAB-BS deployment technology to dynamically adjust the optimal communication parameters according to user requirements and thus enhance the QoS of the deployed UAV-BS.


\begin{figure}[!t]
	\centering
	\includegraphics[width=0.45\textwidth]{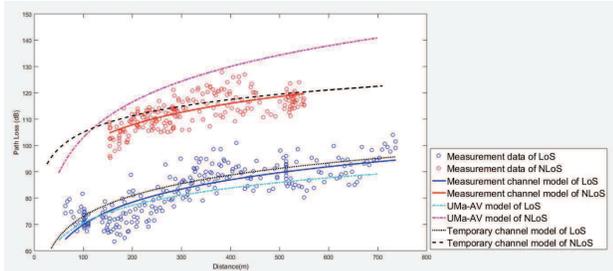}
	\caption{The path-loss result of the constructed channel data model.}
	\label{fig:channel_reconstruction_result}
	\vspace{-12pt}
\end{figure}

In the simulation, we consider the UMa-AV scenario and use the channel model given by 3GPP TR 36.777~\cite{3gpp.36.777} as the baseline. According to Fig.~\ref{fig:channel_reconstruction_result}, the constructed temporary channel model by the proposed online learning approach can be much more close to the measured real channel data. It means that the proposed online learning approach predicts the channel characteristics more accurate than the 3GPP statistical channel model does.

%

\subsection{Data-Driven Placement of UAV-BSs for Arbitrary Crowds}
\begin{figure}[!t]
	\centering
	\includegraphics[width=0.42\textwidth]{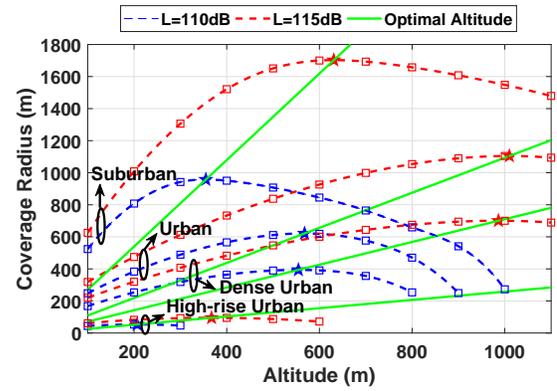}
	\vspace{-5pt}
	\caption{The relationship between the altitude and the coverage of a UAV-BS.}
	\label{fig:coverage_altitude}
	\vspace{-10pt}
\end{figure}

For UAV-BS placement issue, we consider the case of deploying single UAV-BS and focus on downlink communications and propose an on-demand 3D placement~\cite{8642333} which can find the appropriate altitude and horizontal location with the minimum coverage to serve the maximum number of arbitrarily distributed UEs. In general, the decision process of 3D UAV placement is divided into two phases: 1) the altitude optimization of the UAV-BS and 2) the horizontal deployment optimization of the UAV-BS. The relationship between the altitude of the UAV-BS and the maximum coverage is shown in Fig.~\ref{fig:coverage_altitude} and has been discussed in~\cite{6863654}. In stage 1, the system can find the optimal altitude and the maximum available coverage of the UAV-BS. The system then uses the obtained maximum available coverage as a limitation in stage 2. In stage 2, the system uses the proposed on-demand strategy to dynamically deploy UAV-BSs and provides the communication services. 

\begin{figure}[!t]
	\centering
	\includegraphics[width=0.42\textwidth]{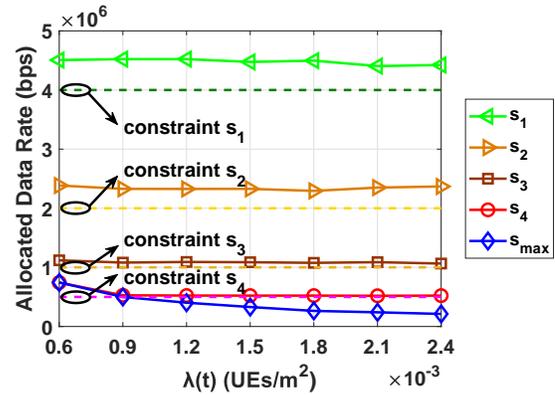}
	\vspace{-5pt}
	\caption{The allocated data rate for different demands in different UE density~\cite{8642333}.}
	\label{fig:allocated_datarate}
	\vspace{-12pt}
\end{figure}

For cellular operators, they can pre-define their own configurations including the parameters of different communication quality requirements, such as data rate, signal-to-interference-plus-noise ratio (SINR), and latency. Fig.~\ref{fig:allocated_datarate} shows that the proposed on-demand 3D placement can always guarantee that the allocated data rate of each UE is higher than the lower bound of each UE demand level. By contrast, with the policy of using the case of $s_{\max}$, a UAV-BS can cover the maximum number of UEs, but such a way leads spectrum contention problem and makes the allocated data rate of each UE intolerable as the UE density increases.

\subsection{The Integration of WiFi Front-haul and WiGig Back-haul Networks}
To provide stable UAV-BS service to UEs, the back-haul transmission of each UAV-BS must be guaranteed. We thus carry out the preliminary integrated development of the WiFi and the millimeter wave (WiGig) communications. WiFi is responsible for providing front-haul connections to UEs and WiGig is responsible for guarantee the high-speed back-haul connections. In order to let terminal devices and drones communicate with each other, we use the \emph{Robot Operating System} (ROS), an open source robot development platform that can support multiple functions required by robots. In the integration of millimeter wave transmission, we chose the USB WiGig network card using Tensorcom chip and connect this card to the development board of the UAV-BS. We integrate a WiFi access point (AP) on the UAV-BS with two antennas and the equipped WiFi chip is Intel Corporation Wireless-AC 3165. In order to achieve high-speed transmission, the interface of the connection must be USB 3.0, so we chose UP Squared development board as the edge system of the UAV-BS, which is responsible for connecting the WiFi and WiGig networks. 

\begin{figure}[!t]
	\centering
	\includegraphics[width=0.35\textwidth]{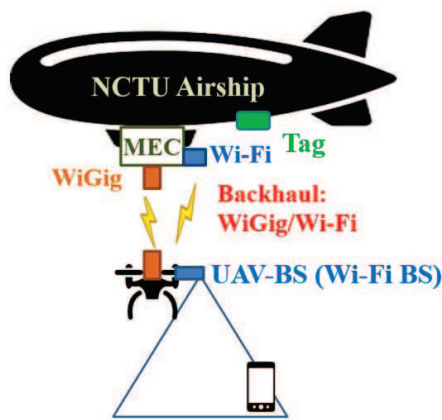}
	\vspace{-5pt}
	\caption{The architecture of the network integration using WiFi and WiGig.}
	\label{fig:integration_architecture}
	\vspace{-5pt}
\end{figure}

The network integration is mainly divided into two stages. First, testing development board to meet the needs of WiGig high-speed transmission. In the second phase, to make the WiGig client on the UAV-BS automatically connect to the base station for high-speed data unloading, thereby adjusting parameters and setting optimization performance. The platform we constructed is shown in Fig.~\ref{fig:integration_architecture}. The latest test results presents that the proposed network integration guarantees that WiGig can achieve the highest throughput of about 950 Mbps. The round trip time (RTT) of accessing the DNS server can also be reduced to less than 5 ms (low latency). In the future, we are going to discuss and test the influence of equipped antenna angle and position on the network throughput.

\subsection{Innovative Design and Development of UAV}
\begin{figure}[!t]
	\centering
	\includegraphics[width=0.4\textwidth]{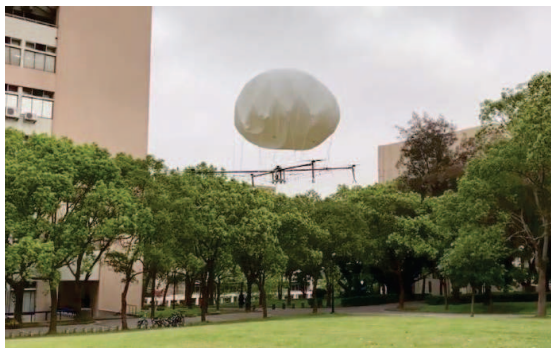}
	\caption{The flight test of the developed new UAV (1st generation)~\cite{20190016461A1}.}
	\label{fig:drone_cruiser}
	\vspace{-10pt}
\end{figure}

After completing the preliminary design of the UAV-BS, in order to systematically solve the possible remaining problems, we first conducted experiments on the drone parts and then confirmed the stability of the flying control system. After completing the initial structural design and electromechanical integration, we conducted a flight test, as shown in Fig.~\ref{fig:drone_cruiser}. The use of Helium balloon can provide the air buoyancy for reducing about 40\% effect of gravity. According to the result of stability test, we have found that the design of 1st-generation UAV's body structure needs improvement. To improve the hovering time, we develop the UAV with lightweight materials. However, we reduce too much weight of UAV to maintain the stability of hovering.

We have designed the 2nd-generation UAV with the consideration of the above issue. The prototype of the 2nd-generation UAV is under construction. After solving the issue of structural design, we are going to develop and apply some machine learning techniques to the control system on our new UAV for improving the stability of hovering.

%


\section{Conclusion}
\label{sec:conclusion}
The goal of this paper is to find a way to 5G-beyond or 6G wireless communications technology by integrating the UAV 3D communications. We have introduced some important issues and proposed some solutions for these issues. We also developed a preliminary version of the simulation platform to help the cellular operators or new service providers to obtain the recommendation of UAV-BS placement. In the future, we will not only keep going on solving the problems of 3D wireless communication and networking but also focus on the other UAV-enabled emerging services and applications.

\bibliographystyle{IEEEtran}
\bibliography{IEEEabrv,reference}

\end{document}